\documentclass[10pt,journal]{IEEEtran}            

\IEEEoverridecommandlockouts                              
\usepackage[utf8]{inputenc}
\usepackage[english]{babel}
\usepackage[normalem]{ulem}
\usepackage{amsmath}
\usepackage{amssymb}
\usepackage{amsfonts}
\usepackage{fouriernc}
\usepackage{algorithm}
\usepackage{algorithmic}
\usepackage{graphicx}
\usepackage{subfig}
\usepackage{float}

\newcommand{\bb}[1]{\mathbb{#1}}
\newcommand{\ca}[1]{\mathcal{#1}}

\usepackage{hyperref}
\newtheorem{thm}{Theorem}[section]
\newtheorem{lm}{Lemma}[section]

\newtheorem{df}{Definition}[section]

\newtheorem{rem}{Remark}[section]
\newtheorem{pr}{Proposition}[section]
\newtheorem{eg}{Example}[section]
\newtheorem{ass}{Assumption}[section]

\usepackage[utf8]{inputenc}
\usepackage[english]{babel}
 
\usepackage[usenames, dvipsnames]{color}
\definecolor{red}{rgb}{1,0.2,0.2}

\usepackage{amssymb}
\title{Notions of Centralized and Decentralized Opacity in Linear Systems$^{*,+}$}
\author{Bhaskar Ramasubramanian$^{1}$, Rance Cleaveland$^{2}$, and Steven I. Marcus$^{1}$
\thanks{$^*$Work supported by the NSF under Grants $CNS-1446665$ and $CMMI-1362303$, and by the AFOSR under Grant $FA9550-15-10050$.}
\thanks{$^+$Preliminary versions of this work was presented in parts at the IEEE American Control Conference, $2016$ \cite{bhaskar2016opacity}, and at the $54^{th}$ Annual Allerton Conference on Communication, Control, and Computing, $2016$ \cite{bhaskar2016decentopacity}.}
\thanks{$^{1}$Department of Electrical and Computer Engineering, and Institute for Systems Research,
University of Maryland, College Park, MD 20742, USA. 
        {\tt\small \{rbhaskar, marcus\}@umd.edu}}%
\thanks{$^{2}$Department of Computer Science, and Institute for Systems Research,
University of Maryland, College Park, MD 20742, USA. 
        {\tt\small rance@cs.umd.edu}}%
}
\date{}
\begin{document}
\maketitle
\begin{abstract} 

We formulate notions of opacity for cyberphysical systems modeled as discrete-time linear time-invariant systems. 
A set of secret states is $k$-ISO with respect to a set of nonsecret states if, starting from these sets at time $0$, the outputs at time $k$ are indistinguishable to an adversarial observer. 
Necessary and sufficient conditions to ensure that a secret specification is $k$-ISO are established in terms of sets of reachable states. 
We also show how to adapt techniques for computing under-approximations and over-approximations of the set of reachable states of dynamical systems in order to soundly approximate k-ISO. 
Further, we provide a condition for output controllability, if $k$-ISO holds, and show that the converse holds under an additional assumption. 

We extend the theory of opacity for single-adversary systems to the case of multiple adversaries and develop several notions of decentralized opacity. 
We study the following scenarios: i) the presence or lack of a centralized coordinator, and ii) the presence or absence of collusion among adversaries. 
In the case of colluding adversaries, we derive a condition for nonopacity that depends on the structure of the directed graph representing the communication between adversaries. 

Finally, we relax the condition that the outputs be indistinguishable and define a notion of $\epsilon$-opacity, and also provide an extension to the case of nonlinear systems. 
\end{abstract}
\section{Introduction} \label{Introduction}

Cyberphysical systems (CPSs) are complex systems in which the functioning of the physical system is governed by computers that communicate instructions and operational protocols.
This is often carried out over a network, which may be wired or wireless, indicating that computational resources and bandwidth could also affect the operation of the CPS. 
CPSs are ubiquitous. 
Examples include power systems, water distribution networks, and on a smaller scale, medical devices and home control systems \cite{baheti2011cyber}. 
While computer-controlled systems allow for the better integration of sensors, actuators, and algorithms, the sharing of information among devices and across geographies makes the system vulnerable to cyber-attacks. 
Such a cyber-attack could be carried out on the physical system, on the computer(s) controlling the system, or on the communication links between the system and the computer. 
Thus, significant material damage can be caused by an attacker who is able to gain access to the system remotely, and such attacks will often have the consequence of causing widespread disruption to everyday life. 
A recent example was an attack on the power grid in Ukraine carried out in December $2015$, where attackers remotely gained access to circuit breakers which brought several substations offline. 
They also remotely disabled backup power supplies, and flooded call centers with fake calls, to prevent affected customers from reporting complaints. 
This left more than $200,000$ people without electricity for several hours. 
The possible impact of a similar attack on the United States power grid is examined in \cite{sullivan2017cyber}. 
Several other instances of attacks on CPSs have been documented, inlcuding in \cite{slay2008lessons}, \cite{farwell2011stuxnet}. 
A compilation of potential challenges in securing control systems is tabled in \cite{cardenas2008research}. 

The starting point for any attack is the collection of information by the attacker from the system that can then be used by the attacker to compromise the system. 
One way of achieving this goal could be for the attacker to focus on the flow of information from the CPS to itself \cite{schneider1996csp, focardi1994taxonomy}. 
Thus, information critical to nominal operation should be safeguarded in a well designed system. 
This motivation has led researchers to develop methods to analyze how \emph{opaque} the system behavior is to an adversary. 
Opacity is a property that captures whether an intruder, modeled as an adversarial observer, can infer a `secret' of a system based on its observation of the system behavior. 
The current state of the art in this area studies opacity within the framework of discrete event systems (DESs) described by regular languages \cite{badouel2007concurrent, saboori2007notions}. 
Techniques from supervisory control can be used to enforce opacity on a system \cite{dubreil2010supervisory,saboori2012opacity}. 
In other words, a controller can be designed to disable actions that lead to the leaking of the secret. 

\textbf{Motivation}: Although the theory of opacity for DESs is quite rich, a shortcoming is that it only studies the case when the states are discrete (like in a DES). 
In many practical systems, it is common for the system variables to take values in a continuous domain. 
This is indeed the case in CPSs like power systems and water distribution networks. 
This paper considers CPSs modeled as a discrete-time linear time-invariant (DT-LTI) system \cite{pasqualetti2015control} (thus, while time steps are discrete, the state, control, and output variables are real valued). 
We use tools from control theory to study opacity for such systems. 
To the best of our knowledge, this paper is the first to present a treatment of opacity for continuous-state systems. 

\textbf{Contributions}: We define a notion of opacity at a time $k$ called \emph{$k$-initial state opacity ($k$-ISO)}. 
A set of secret states is said to be $k$-ISO with respect to a set of nonsecret states if the outputs at time $k$ of every trajectory starting from the set of secret states cannot be distinguished from the output at time $k$ of some trajectory starting from the set of nonsecret states. 
Necessary and sufficient conditions to achieve $k$-ISO are presented in terms of sets of reachable states. 
To overcome the challenge of exactly computing these reachable sets, we present a characterization of our notion of opacity in terms of overapproximations and underapproximations of these sets. 
Opacity of a given DT-LTI system is shown to be equivalent to the output controllability of a system obeying the same dynamics, but with different initial conditions. 
Furthermore, we examine $k$-ISO under unions and intersections of sets of states. 

We extend our treatment to the case of multiple adversarial observers and define several notions of decentralized opacity. 
These notions of decentralized opacity will depend on whether there is a centralized coordinator or not, and the presence or absence of collusion among the adversaries. 
We establish conditions for decentralized opacity in terms of sets of reachable states. 
In the case of colluding adversaries, we derive a condition for \emph{nonopacity} in terms of the structure of the communication graph. 
Finally, we study two extensions: \emph{i)} we relax the necessity of indistinguishability of outputs in the definition of $k$-ISO, and define a notion of $\epsilon$-opacity; and \emph{ii)} we characterize opacity for discrete-time nonlinear systems.

This paper builds on the preliminary work presented in \cite{bhaskar2016opacity} and \cite{bhaskar2016decentopacity} in the following ways: \emph{i)} we present the results for the single and multiple adversary cases in greater detail with complete proofs; \emph{ii)} new results on the relationship between $k-$ISO and approximations of reachable sets of states are given, as are new results on $k-$ISO for nonlinear systems.
%
%
\subsection{Related Work}

Opacity was first presented as a tool to study cryptographic protocols in \cite{mazare2004using}. 
The intruder was modeled as a passive observer who could read messages exchanged between two parties, but could not modify, block, or send a message. 
The aim of the parties was to exchange secret information without making it accessible to the intruder. 
A theory of supervisory control for DESs represented by finite state automata (FSA) and regular languages was formulated in \cite{ramadge1987supervisory,ramadge1989control}. 
This framework spawned research in many areas including fault diagnosis \cite{sampath1996failure}, hybrid systems \cite{van2000introduction}, and robotics \cite{burridge1999sequential}. 

DESs were used to study opacity in \cite{badouel2007concurrent}, which assumed multiple intruders with different observation capabilities. 
Under the assumption that the supervisor could control all events, it was shown that there exists an optimal control that enforced opacity. 
Verification of the opacity of a secret specified as a language was presented in \cite{dubreil2010supervisory,cassez2012synthesis}, while \cite{saboori2007notions, saboori2008verification, saboori2012verification} studied the same for secrets specified as states. 
Language and state based notions of opacity were shown to be equivalent in \cite{wu2013comparative}, where algorithms to transform one notion of opacity to the other were presented. 
A notion of joint opacity was also proposed in this paper, in which a system was observed by multiple adversarial observers who shared their observations with a coordinator, which then verified opacity.

Opacity was compared with detectability and diagnosability of DESs, and other privacy properties like secrecy and anonymity in \cite{lin2011opacity}. 
A subsequent paper \cite{paoli2012decentralized} defined opacity for DESs in a decentralized framework with multiple adversaries, each carrying out its own observation of the system. 
The authors of \cite{ben2009opaque} characterized language-based notions of opacity under unions and intersections. 
They demonstrated the existence of supremal and minimal opaque sublanguages and superlanguages. 

Enforcement of opacity using techniques from supervisory control was studied in \cite{dubreil2010supervisory, saboori2012opacity}. 
The authors of \cite{wu2014synthesis} formulated an alternate method of opacity enforcement using insertion functions, which are entities that modify the output behavior of the system in order to keep a secret.  
The model-checking and verification of notions of opacity at run time in online setups was presented in \cite{falcone2015enforcement}. 
A scheme for the verification of opacity in DESs using two-way observers was proposed in \cite{yin2017new}. 
This enabled a unified framework to verify multiple notions of opacity.

There is a large body of literature focused on developing techniques to compute overapproximations and underapproximations of sets of reachable states. 
These will usually depend on how the initial set of states is specified, including support functions \cite{girard2008efficient}, zonotopes \cite{girard2006efficient}, and ellipsoids \cite{kurzhanskiy2007ellipsoidal}. 
A method to compute overapproximations of reachable sets of states for linear systems with uncertain, time-varying parameters and inputs was presented in \cite{althoff2011reachable}. 
The reader is referred to \cite{maler2008computing} for a succinct presentation of some of the techniques used in computing reachable sets. 
\subsection{Outline of Paper}

The remainder of this paper is organized as follows: Section \ref{OpacityDES} provides an introduction to opacity as studied for DESs. 
Section \ref{OpacityLTI} presents a new formulation of opacity for DT-LTI systems. 
In Section \ref{OpacityReachableSets}, we present necessary and sufficient conditions that would establish our notion of opacity in terms of reachable sets of states. 
Section \ref{UandI} studies opacity under unions and intersections. 
Section \ref{ReachOpaApprox} uses overapproximations and underapproximations of the reachable set of states to characterize the notions of opacity presented in Section \ref{OpacityLTI}. 
Opacity is analyzed in terms of the notion of output controllability in Section \ref{OpacityOutputCont}. 
We then extend our framework to the case of multiple adversaries in Section \ref{OpacityDecent}, where we present more than one notion of decentralized opacity. 
A notion of opacity where a somewhat strict assumption is relaxed is presented in Section \ref{OpacityEpsilon}, and an extension to the case of nonlinear systems is proposed in Section \ref{OpacityNonLin}. 
We conclude by presenting future directions of research in Section \ref{Conclusion}. 
\section{Opacity for Discrete Event Systems} \label{OpacityDES}

This section gives a brief introduction to the notions of opacity studied for discrete event systems. 
The DES is typically modeled as a finite state automaton, and the secret could be specified as a subset of states or a sublanguage of this automaton. 
%

Let $\Sigma$ be an alphabet, and let $\Sigma^*$ be the set of all strings of elements from $\Sigma$ of finite length, including the empty string $\epsilon$. 
A language $\mathcal{L}$ is a subset of $\Sigma^*$. 
Let $G=(X,\Sigma,f,X_0)$ be an finite state automaton, where $X$ is a nonempty set of states, $X_0 \subseteq X$ is a nonempty set of initial states, and $\Sigma$ represents the set of events. 
$f:X \times \Sigma \rightarrow X$ is the (partial) state transition function: given $x,y \in X$ and $\sigma \in \Sigma$, $f(x,\sigma)=y$ if the execution of $\sigma$ from $x$ takes the system to $y$. 
We write $f(x,\sigma)!$ if $f(x,\sigma)$ is a valid transition. 
The transition function is extended to $f:X \times \Sigma^* \rightarrow X$ in the usual recursive way: 
\begin{align}
f(x,\epsilon)&:=x \nonumber \\
f(x,se)&:=f(f(x,s),e) \text{ for } s \in \Sigma^*, e \in \Sigma. \nonumber
\end{align}
The language generated by $G$ is $\mathcal{L}(G):= \{s \in \Sigma^* : f(x,s)!\}$, and describes all possible trajectories of the system.  

Let $P: \Sigma^* \rightarrow \Sigma^*$ be a projection map. 
Then, if a string of events $s$ occurs in the system, an external agent would see $P(s)$. 
$P$ can be extended from strings to languages as follows: for languages $L, J \subseteq \Sigma^*$, define
\begin{align}
P(L)&=\{t \in \Sigma^* : (\exists s \in L)t = P(s) \} \nonumber \\
P^{-1}(J)&=\{t \in \Sigma^* : P(t) \in J \} \nonumber 
\end{align}
%

A secret specification (states, or language) is \emph{opaque} with respect to a nonsecret specification if every secret execution (will be made clear subsequently) is indistinguishable from a nonsecret execution. 
The notion of opacity under consideration will depend on how the secret (sublanguage, set of initial states, or set of current states) is specified. 
Let $K_1$ and $K_2$ be sublanguages of $\mathcal{L}(G)$.  

\begin{df}[Language-based Opacity for DES]
$K_1$ is \emph{language based opaque (LBO)} with
respect to $K_2$ and $P$ if for every trajectory in $K_1$, there exists a trajectory
in $K_2$ that `looks' the same under $P$, i.e. $K_1 \subseteq P^{-1}(P(K_2))$.
\end{df}
%
%
\begin{df}[Initial State Opacity for DESs]
Given $G$ with $X_s,X_{ns} \subseteq X_0$, and $P$, $X_s$ is \emph{initial state
  opaque (ISO)} with respect to $X_{ns}$ and $P$ if for every $i
\in X_s$ and every $t \in L(G,i)$ such that $f(i,t)$ is defined, there
exists $j \in X_{ns}$ and $t' \in L(G,j)$ such that $f(j,t')$ is
defined and $P(t)=P(t')$.
\end{df}
%

There are other definitions of opacity that are based on the current state of the DES, rather than the initial state, as defined above. 
These state based and language based definitions are essentially equivalent, since it has been shown that there exist polynomial time algorithms that relate any pair of the notions of opacity \cite{wu2013comparative}.
The reader is referred to \cite{saboori2007notions, wu2013comparative,lin2011opacity} for a more detailed exposition on opacity for discrete event systems. 
%
%
\begin{eg}{\cite{saboori2008verification}}
\begin{figure}[h]
 \centering
  \includegraphics[width=2 in]{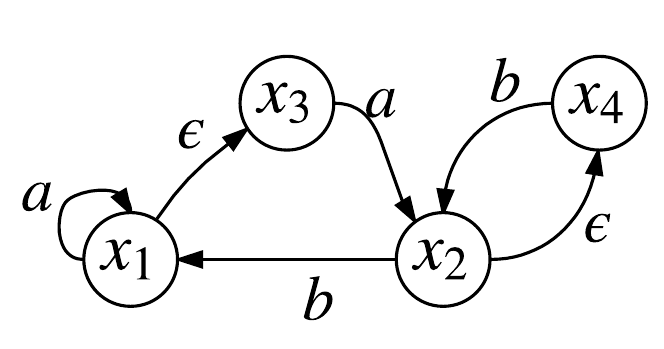} 
\caption{Initial State Opacity} \label{DESISO} 
\end{figure}
Consider the FSA $G$ in Figure \ref{DESISO}, with observable events given by $\Sigma_o:=\{a,b\}$. Assume that $X_s=\{x_3\}$ and $X_{ns}=X \setminus X_s$. 
Then, $X_s$ is ISO w.r.t. $X_{ns}$ because for every string $s$ starting from $x_3$, there is another string $\epsilon s$ starting from $x_1$, that looks the same. 
However, $X_s=\{x_1\}$ will not be ISO w.r.t. $X_{ns} = X \setminus X_s$. 
In this case, whenever an intruder sees the string $aa$, it will be sure that the system started from $X_s$ (since no other initial state can generate a string that appears the same as $aa$). 
\end{eg}
\section{Opacity for Linear Time-Invariant Systems} \label{OpacityLTI}

Although the presentation of opacity for DESs is well-motivated and elegant, a shortcoming of the framework is that it only addresses the case when the states of the system are discrete. 
The states in CPSs like power systems and water networks are typically real valued.  
It is this gap that we seek to bridge in this paper, by formulating notions of opacity for continuous state systems. 
The system is modeled as a discrete-time linear time-invariant system. 
Therefore, while the time steps are discrete, the states, input, and output variables are real valued. 

We define a notion of opacity for such systems, called $k$-initial state opacity. 
We present conditions that will establish $k$-ISO of a set of `secret' initial states in terms of sets of reachable states, and in relation to the notion of output controllability. 
Moreover, we will develop a characterization of $k$-ISO in terms of overapproximations and underapproximations of the reachable sets of states. 

Consider the system: 
\begin{align}
x(t+1)&=Ax(t)+Bu(t) \nonumber \\
x(0)&=x_0 \in X_0 \nonumber \\
y(t)&=Cx(t) \label{LTISys}
\end{align}
where $x \in \mathbb{R}^n, u \in \mathbb{R}^m, y \in \mathbb{R}^p$, and $A,B,C$ are matrices of appropriate dimensions containing real entries. 

Let $\mathcal{K}$ be a set of positive integers, corresponding to instants of time at which the adversary makes an observation of the system. 
The subscript $s$ ($ns$), when appended to the states, inputs, and outputs, will correspond to trajectories that start from the set of initial secret (nonsecret) states. 
The adversary is assumed to have knowledge of the initial sets of secret and nonsecret states, $X_s$ and $X_{ns}$, the system model $(A,B)$, and its own observation map $C$. 
Further, we assume that it has unlimited computing power, in that it will be able to compute the sets of reachable states at time $k$. 
Its goal is to deduce, on the basis of observing the system at times $k \in \ca{K}$, whether the system started from a state in $X_s$ or not. 

\begin{df}[Strong $k-$Initial State Opacity] \label{OpacityDefn}
For the system (\ref{LTISys}), given $X_s, X_{ns} \subset X_0$ and $k \in \ca{K}$, $X_s$ is \emph{strongly $k$-initial state opaque ($k$-ISO)} with respect to $X_{ns}$ if for every $x_s(0) \in X_s$ and for every sequence of admissible controls $u_s(0),\dots,u_s(k-1)$, there exist an $x_{ns}(0) \in X_{ns}$ and a sequence of admissible controls $u_{ns}(0),\dots,u_{ns}(k-1)$ such that $y_s(k)=y_{ns}(k)$. 

$X_s$ is \emph{strongly $\ca{K}$-ISO} with respect to $X_{ns}$ if $X_s$ is strongly $k$-ISO with respect to $X_{ns}$ for all $k \in \ca{K}$. 
\end{df}

This means that starting from any secret state and applying any sequence of $k$ admissible controls (corresponding to the instants the adversary makes an observation), the system will reach a state whose observation to the adversary will be indistinguishable from the observation of a state that can be reached by the application of an admissible control sequence of length $k$, starting from some nonsecret state. 
Without loss of generality, the sets $X_s$ and $X_{ns}$ are disjoint. 
Definition \ref{OpacityDefn} is independent of whether these sets partition the entire state space of the system.  
While this notion calls for every state in the set of initial secret states to be indistinguishable (after some time $k$) from some state in the initial set of nonsecret states, the following definition relaxes this requirement. 

\begin{df}[Weak $k-$Initial State Opacity]
For the system (\ref{LTISys}), given $X_s, X_{ns} \subset X_0$ and $k \in \ca{K}$, $X_s$ is \emph{weakly $k$-ISO} with respect to $X_{ns}$ if for some $x_s(0) \in X_s$ and for some sequence of admissible controls $u_s(0),\dots,u_s(k-1)$, there exist an $x_{ns}(0) \in X_{ns}$ and a sequence of admissible controls $u_{ns}(0),\dots,u_{ns}(k-1)$ such that $y_s(k)=y_{ns}(k)$. 

$X_s$ is \emph{weakly $\ca{K}$-ISO} with respect to $X_{ns}$ if $X_s$ is weakly $k$-ISO with respect to $X_{ns}$ for all $k \in \ca{K}$. 
\end{df}

These definitions of opacity for LTI systems is different from familiar definitions of observability. 
The observability problem aims to determine the initial state $x(0)$, given the entire output and control histories. 
Here, however, an adversary aims to determine $x(0)$ via access to only snapshots of the output and the set of possible controls. 
This supports reasoning about adversaries with limited observational capabilities.
It could also be the case that an adversary might not want to reveal its presence, or not have the resources to continuously monitor the system.
%

Our formulation is also different from definitions of opacity in the DES literature. 
In those cases, the observation of the entire secret trajectory must coincide with that of a nonsecret trajectory. 
We only need that the secret and nonsecret outputs at time $k$ coincide. 
$k$-ISO also differs from the notion of $k$-step opacity proposed in \cite{saboori2011verification}. 
In their formulation, $k$-step opacity is achieved when the adversary does not know if the system entered a secret state in $k$ previous steps. 
We require that the ambiguity exist only at time $k$. 
An additional requirement to our conditions for $k$-ISO will also establish $k$-step opacity.

Finally, $k$-ISO is also different from the notion of bisimulation relations between dynamical systems \cite{van2004equivalence}. 
Bisimulation relations typically verify the `equality' of two systems governed by different dynamics. 
In our framework, however, we try to identify equivalence classes of outputs at time $k$. 
Opacity is deemed to have been achieved if the system starting from two disjoint sets of states at time $0$ reaches the same equivalence class of outputs at time $k$. 

\begin{eg}
\begin{figure}
 \centering
  \includegraphics[width=3 in]{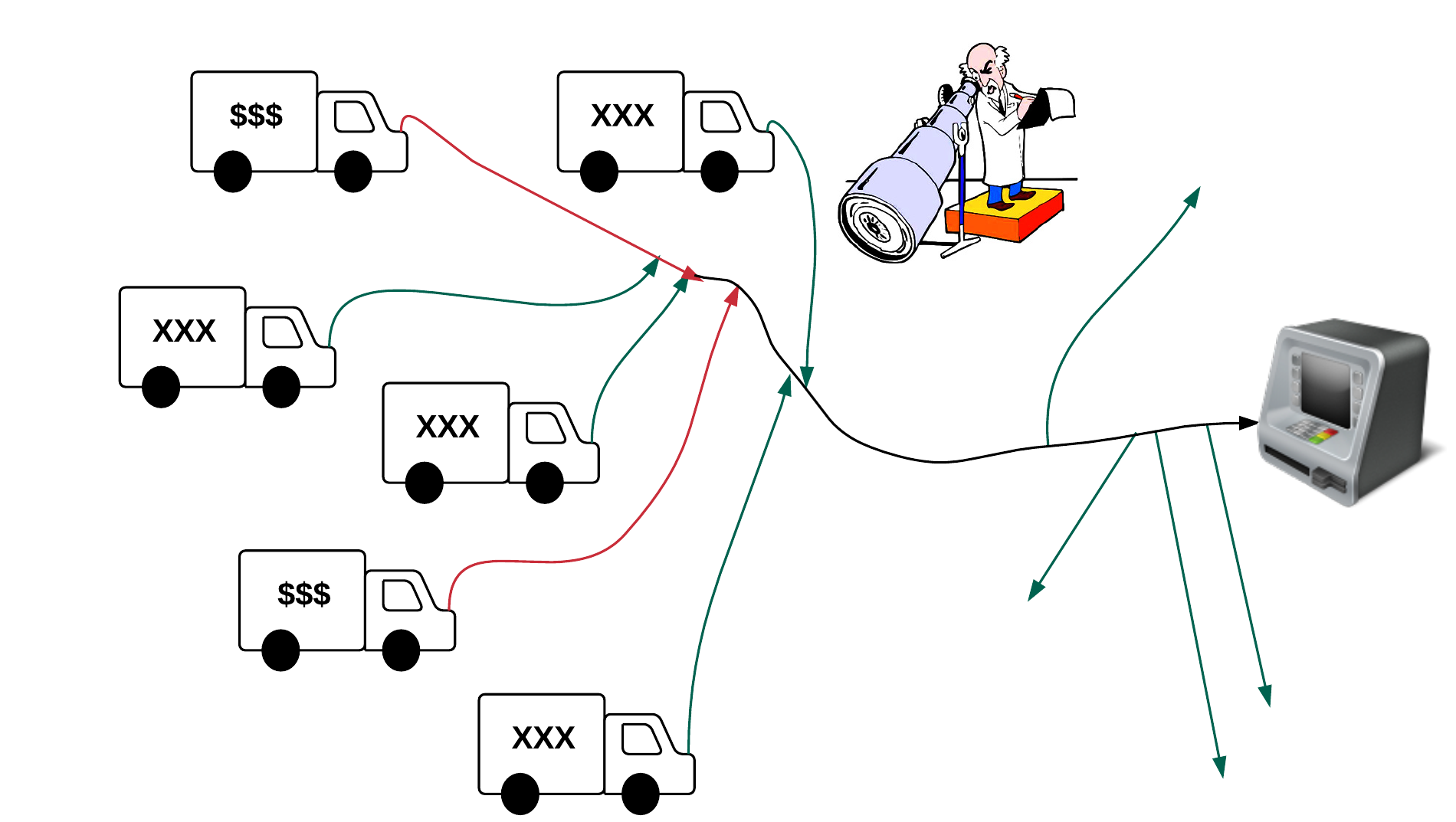} 
\caption{ATM Money Transfer} \label{OpaEg} 
\end{figure}
Figure \ref{OpaEg} illustrates the problem of a bank needing to transfer money from its office to an ATM machine. 
One way for the bank to do this in a `secure' manner would be to equip the truck with the best defenses that money can buy. 
However, this might not be a cost-effective solution, since customizations might be very expensive, and need to be continuously updated to stay ahead of potential attackers. 
An alternative approach would be for the bank to deploy several identical trucks, only some of which carry money. 
This is a reasonable strategy for the bank to adopt, under the assumption that the cost of carrying out an attack on a truck is very high. 

The motion of a truck can be represented by a system whose states are the position and velocity of the truck,  
and input is the acceleration. 
Then, assuming unit mass, and unit sampling interval, the system model is:
\begin{align*}
\begin{bmatrix} p(k+1)\\v(k+1) \end{bmatrix}&=\begin{bmatrix}1&1\\0&1\end{bmatrix}\begin{bmatrix} p(k)\\v(k) \end{bmatrix}+\begin{bmatrix} 0.5\\1 \end{bmatrix}a(k)\\
y(k)&=\begin{bmatrix} 1&0 \end{bmatrix}\begin{bmatrix} p(k)\\v(k) \end{bmatrix}
\end{align*}

The position of the truck at a time $k$ is given by: 
\begin{align*}
p(k)&=p(0)+kv(0)+\sum_{i=0}^{k-1}(k-i-0.5)a(i)
\end{align*}

Let the locations at which money is loaded (denoted $\$\$\$$) comprise the set of secret states ($X_s$), and the initial locations of the other trucks (denoted $XXX$) comprise the set of nonsecret states ($X_{ns}$). 
Then, if an adversary observes $p(k)$ at some time $k$, $X_s$ will be $k$-ISO with respect to $X_{ns}$ if it cannot determine whether the truck started from a $\$\$\$$ location or an $XXX$ location. 
That is, for every location (at time $0$) at which money was loaded, there is a corresponding location (at time $0$) where there was no money loaded such that the positions of trucks which start from these locations are the same at time $k$. 
\end{eg}

\section{Opacity and Reachable Sets of States} \label{OpacityReachableSets}

The adversary has complete knowledge of the system model, and the sets of initial secret and nonsecret states. 
However, it does not know the exact control sequence applied in the time interval $[0,k]$; it only has knowledge of the sets of allowed inputs that can be applied. 
In this light, a possible means of checking that opacity holds is by relating it to reachability. 
In this section, we present necessary and sufficient conditions to establish $k$-ISO in terms of sets of reachable states of the system. 

Let $U_{s}^k:=\{u_{s}(0),\dots,u_{s}(k-1)\}$ be a sequence of $k$ controls starting from an initial secret state. 
We analogously define $U_{ns}^k:=\{u_{ns}(0),\dots,u_{ns}(k-1)\}$. 
Let $X_s(k)$ and $X_{ns}(k)$ denote the sets of states reachable in $k$ steps, starting at time $0$ from \emph{nonempty} sets $X_s$ and $X_{ns}$ respectively. 
That is, 
\begin{align}
X_s(k) &= \bigcup _{x_0 \in X_s} \bigcup _{U_{s}^k}\{x: x(i+1)=Ax(i)+Bu(i), \forall i <k \} \label{secreach} \\
X_{ns}(k)&= \bigcup _{x_0 \in X_{ns}} \bigcup_{U_{ns}^k}\{x: x(i+1)=Ax(i)+Bu(i), \forall i <k \} \label{nonsecreach}
\end{align}

The set of outputs at time $k$ can be represented by
\begin{align}
CX(k)&:=\{y: y = Cx, x \in X(k)\}
\end{align}

\begin{thm} \label{OpaIff}
The following hold: 
\begin{enumerate}
\item $X_s$ is strongly $k$-ISO with respect to $X_{ns}$ if and only if $CX_s(k) \subseteq
CX_{ns}(k)$. 

\item $X_s$ is strongly $\ca{K}$-ISO with respect to $X_{ns}$ if and only if $CX_s(k) \subseteq
CX_{ns}(k)$ for all $k \in \ca{K}$.
\end{enumerate}
\end{thm}

\begin{IEEEproof}
First, let strong $k$-ISO hold. 
Then, for all $x_s(0) \in X_s$, and all $\{u_s(\cdot)\}_0^{k-1}$, there exist $x_{ns}(0) \in X_{ns}$ and $\{u_{ns}(\cdot)\}_0^{k-1}$ such that $y_s(k)=y_{ns}(k)$. 
Now, starting from $X_s$ (respectively $X_{ns}$), and applying $k$ admissible controls, one reaches a state in $X_s(k)$ ($X_{ns}(k)$). 
Therefore, $k$-ISO ensures that for every $x_s(k) \in X_s(k)$, there exists $x_{ns}(k) \in X_{ns}(k)$, such that $y_s(k) = y_{ns}(k)$. 
This gives $CX_s(k) \subseteq CX_{ns}(k)$. 

Next, let $CX_s(k) \subseteq CX_{ns}(k)$. 
This means for every $x_s(k) \in X_s(k)$, there exists $x_{ns}(k) \in X_{ns}(k)$, such that $y_s(k) = y_{ns}(k)$. 
Since $X_s(k)$ and $X_{ns}(k)$ are sets of reachable states starting from $X_s$ and $X_{ns}$ respectively, the previous sentence translates to: 
for every $x_s(0) \in X_s$ and every $\{u_s(\cdot)\}_0^{k-1}$, there exist $x_{ns}(0) \in X_{ns}$ and $\{u_{ns}(\cdot)\}_0^{k-1}$ such that $y_s(k)=y_{ns}(k)$. 
This, by definition, is strong $k$-ISO. 

The second statement of the theorem easily follows by extending the above argument to all $k \in \ca{K}$. 
\end{IEEEproof}

\begin{rem}
This result can be easily extended to verify $k$-step opacity if $CX_s(k) \subseteq CX_{ns}(k)$ for all $k \in \ca{K}:=\{m,m-1,\dots,m-k+1\}$ for any positive integer $m>k$. 
\end{rem}

\begin{rem}
$X_s(k) \subseteq X_{ns}(k)$ is only a sufficient condition for $X_s$ to be strongly $k$-ISO with respect to $X_{ns}$. 
To see that this condition is not necessary, let $C=\bigl(\begin{smallmatrix}
1&1&1
\end{smallmatrix} \bigr)$, and $X_s(k) = \bigl(\begin{smallmatrix}
1&0&0
\end{smallmatrix} \bigr)^T$ and $X_{ns}(k) = \bigl(\begin{smallmatrix}
0&1&0
\end{smallmatrix} \bigr)^T$. 
Then, $CX_s(k) = CX_{ns}(k)$, establishing $k-$ISO, even though
$X_s(k) \not \subseteq X_{ns}(k)$. 
\end{rem}

A similar results holds for weak $k-$ISO. 
%
\begin{thm}
The following hold: 
\begin{enumerate}
\item $X_s$ is weakly $k$-ISO with respect to $X_{ns}$ if and only if $CX_s(k) \cap
CX_{ns}(k) \neq \phi$. 
\item $X_s$ is weakly $\ca{K}$-ISO with respect to $X_{ns}$ if and only if $CX_s(k) \cap
CX_{ns}(k) \neq \phi$ for all $k \in \ca{K}$. 
\end{enumerate}
\end{thm}

\begin{rem}
$X_s(k) \cap X_{ns}(k) \neq \phi$ is a sufficient condition for $X_s$ to be weakly $k$-ISO with respect to $X_{ns}$. 
\end{rem}

The next example helps illustrate Theorem \ref{OpaIff}. 

\begin{eg} \label{toyeg}
Let $X_s,X_{ns} \subseteq X_0$ with $X_s=\{ \bigl(\begin{smallmatrix}
1&0&0
\end{smallmatrix} \bigr)^T, \bigl(\begin{smallmatrix}
0&0&1
\end{smallmatrix} \bigr)^T\}$ and $X_{ns}=\mathbb{R}^3 \setminus X_s$. 
Let $A=I_{3 \times 3}$, $B=\bigl(\begin{smallmatrix}
1&1&1
\end{smallmatrix} \bigr)^T$ and $C=\bigl(\begin{smallmatrix}
1&1&1
\end{smallmatrix} \bigr)$. 
The state and output at time $i$ for the dynamics in (\ref{LTISys}) for $x_s(0)=\bigl(\begin{smallmatrix}
1&0&0
\end{smallmatrix} \bigr)^T$ are:
\begin{align}
x_s(i)&= \begin{pmatrix}
1+ \sum_{j=0}^{i-1}u_s(j) &
\sum_{j=0}^{i-1}u_s(j) &
\sum_{j=0}^{i-1}u_s(j) \end{pmatrix}^T\nonumber \\
y_s(i)&=1+3 \sum_{j=0}^{i-1}u_s(j) \label{SecISO}
\end{align}
$x_s(0)= \bigl(\begin{smallmatrix}
0&0&1
\end{smallmatrix} \bigr)^T$ will also give the same $y_s(i)$. 
Now, let $x_{ns}(0)=\bigl(\begin{smallmatrix}
0&1&0
\end{smallmatrix} \bigr)^T$. 
From (\ref{LTISys}), the state and output at time $i$, with initial state $x_{ns}(0)$, are given by: 
\begin{align}
x_{ns}(i)&=\begin{pmatrix}
\sum_{j=0}^{i-1}u_{ns}(j) &
1+\sum_{j=0}^{i-1}u_{ns}(j) &
\sum_{j=0}^{i-1}u_{ns}(j) \end{pmatrix}^T\nonumber \\
y_{ns}(i)&=1+3 \sum_{j=0}^{i-1}u_{ns}(j) \label{NonSecISO}
\end{align}
$x_{ns}(0)=\bigl(\begin{smallmatrix}
0&1&0
\end{smallmatrix} \bigr)^T$ will also work for $x_s(0)= \bigl(\begin{smallmatrix}
0&0&1
\end{smallmatrix} \bigr)^T$.

Comparing equations (\ref{SecISO}) and (\ref{NonSecISO}), $X_s$ will be strongly $k-$ISO w.r.t. $X_{ns}$ if for every admissible control sequence $\{u_s(0),\dots,u_s(k-1)\}$, there is an admissible control sequence $\{u_{ns}(0),\dots,u_{ns}(k-1)\}$ such that: 
\begin{align}
\sum_{j=0}^{k-1}u_{s}(j)&=\sum_{j=0}^{k-1}u_{ns}(j) 
\end{align}
\end{eg}
\section{$k-$ISO Under Set Operations} \label{UandI}

Properties of $k-$ISO are studied under unions and intersections. 
The properties verified will be for strong $k-$ISO, unless otherwise mentioned. 
Let $X$ denote the set of initial states, and $X(k)$ be the set of states reachable in $k$ steps, starting from $X$ at time $0$. 
Throughout this section, $\wedge$ and $\vee$ denote the logical AND and OR operations respectively.

We first study the effect of the set union operation on $k$-ISO. 
In order to do this, we first establish basic results on sets of reachable states and outputs under set union. 
\begin{lm} \label{prop1}
Given sets of initial states $X_1,X_2,\dots \subseteq X$, the reachable set in $k$ steps of their union is equal to the union of the reachable sets in $k$ steps of each set of initial states. That is, 
$(\bigcup_i X_i)(k)=\bigcup_i X_i(k)$. 
\end{lm}
%

\emph{Proof}:
$\begin{aligned}[t]
&x \in (\bigcup_i X_i)(k) \nonumber \\
\Leftrightarrow &\exists x_0 \in (\bigcup_i X_i), \exists \{u(\cdot)\}, \text{ (\ref{LTISys}) holds } \forall i<k, x(k)=x \nonumber \\
\Leftrightarrow &[(\exists x_0 \in X_1 \wedge \exists \{u(\cdot)\}) \text{ s.t. }(x \in X_1(k))] \vee \nonumber \\
&[(\exists x_0 \in X_2 \wedge \exists \{u(\cdot)\}) \text{ s.t. }(x \in X_2(k))] \vee \dots \nonumber \\
\Leftrightarrow &x \in \bigcup_i X_i(k) \nonumber \text{\hspace{47mm}}\blacksquare
\end{aligned}$
%
\begin{lm} \label{prop2}
Given $X_1,X_2,\dots \subseteq X$ and $C: \mathbb{R}^n \rightarrow \mathbb{R}^m$, 
$C(\bigcup_i X_i)(k)=\bigcup_i CX_i(k)$. 
\end{lm}

\emph{Proof}:
$\begin{aligned}[t]
&y \in C(\bigcup_i X_i)(k) \nonumber \\
\Leftrightarrow &\exists x \in (\bigcup_i X_i)(k) \text{ such that } y=Cx \nonumber \\
\Leftrightarrow &\exists x \in \bigcup_i X_i(k) \text{ such that } y=Cx \nonumber \\
\Leftrightarrow &(y=Cx \wedge x \in X_1(k)) \vee (y=Cx \wedge x \in X_2(k)) \vee \dots \nonumber \\
\Leftrightarrow &(y \in CX_1(k)) \vee (y \in CX_2(k)) \vee \dots \nonumber \\
\Leftrightarrow &y \in \bigcup_i CX_i(k) \nonumber  \text{\hspace{48mm}}\blacksquare
\end{aligned}$ 

Lemmas \ref{prop1} and \ref{prop2} are used to study $k$-ISO under set union, as shown in the following two results.
%
\begin{thm} \label{prop3}
$X_{s_i}$ is $k-$ISO with respect to $X_{ns}$ for each $i$ if and only if $\bigcup_i X_{s_i}$ is $k-$ISO with respect to $X_{ns}$.
\end{thm}

\emph{Proof}:
$\begin{aligned}[t]
&X_{s_i} \text{ }k-\text{ISO w.r.t. }X_{ns} \forall i \nonumber \\
\Leftrightarrow &CX_{s_i}(k) \subseteq CX_{ns}(k) \forall i \nonumber \\
\Leftrightarrow &\bigcup_i CX_{s_i}(k) \subseteq CX_{ns}(k) \nonumber \\
\Leftrightarrow &C(\bigcup_i X_{s_i}(k)) \subseteq CX_{ns}(k) \nonumber \\
\Leftrightarrow &\bigcup_i X_{s_i} \text{ is }k- \text{ISO w.r.t. }X_{ns} \nonumber\text{\hspace{28mm}}\blacksquare
\end{aligned} $
\begin{thm} \label{prop4}
$X_{s}$ is $k-$ISO w.r.t. $X_{ns_i}$ for each $i$ if and only if $X_s$ is $k-$ISO w.r.t. $\bigcup_i X_{ns_i}$
\end{thm}

\emph{Proof}:
$\begin{aligned}[t]
&X_{s} \text{ }k-\text{ISO w.r.t. }X_{ns_i} \forall i \nonumber \\
\Leftrightarrow &CX_{s}(k) \subseteq CX_{ns_i}(k) \forall i \nonumber \\
\Leftrightarrow &CX_{s}(k) \subseteq \bigcup_i CX_{ns_i}(k) \nonumber \\
\Leftrightarrow &CX_s(k) \subseteq C(\bigcup_i X_{ns_i}(k)) \nonumber \\
\Leftrightarrow &X_{s} \text{ }k-\text{ISO w.r.t. }\bigcup_i X_{ns_i} \nonumber \text{\hspace{30mm}}\blacksquare
\end{aligned}$ 

We first establish basic results on sets of reachable states and outputs under set intersection, and use these to prove our main results on $k$-ISO under set intersection.
%
\begin{lm} \label{prop5}
Given sets of initial states $X_1,X_2,\dots \subseteq X$, the reachable set in $k$ steps of the intersection of the sets of initial states is contained in the intersection of the reachable sets in $k$ steps of each set of initial states. 
That is,
$(\bigcap_i X_i)(k)\subseteq \bigcap_i X_i(k)$. 
\end{lm}

\emph{Proof}:
$\begin{aligned}[t]
&x \in (\bigcap_i X_i)(k) \nonumber \\
\Rightarrow &\exists x_0 \in (\bigcap_i X_i), \exists \{u(\cdot)\}, \text{ (\ref{LTISys}) holds } \forall i<k, x(k)=x  \nonumber \\
\Rightarrow &[(\exists x_0 \in X_1 \wedge \exists \{u(\cdot)\}) \text{ s.t. }(x \in X_1(k))] \wedge \nonumber \\
&[(\exists x_0 \in X_2 \wedge \exists \{u(\cdot)\}) \text{ s.t. }(x \in X_2(k))] \wedge \dots \nonumber \\
\Leftrightarrow &x \in \bigcap_i X_i(k) \nonumber \text{\hspace{50mm}}\blacksquare
\end{aligned}$
\begin{lm} \label{prop6}
Given $X_1,X_2,\dots \subseteq X$ and $C: \mathbb{R}^n \rightarrow \mathbb{R}^m$, 
$C(\bigcap_i X_i)(k)\subseteq \bigcap_i CX_i(k)$. 
\end{lm}

\emph{Proof}:
$\begin{aligned}[t]
&y \in C(\bigcap_i X_i)(k) \nonumber \\
\Leftrightarrow &\exists x \in (\bigcap_i X_i)(k) \text{ such that } y=Cx \nonumber \\
\Rightarrow &\exists x \in \bigcap_i X_i(k) \text{ such that } y=Cx \nonumber \\
\Leftrightarrow &(y=Cx \wedge x \in X_1(k)) \wedge (y=Cx \wedge x \in X_2(k)) \wedge \dots \nonumber \\
\Leftrightarrow &(y \in CX_1(k)) \wedge (y \in CX_2(k)) \wedge \dots \nonumber \\
\Leftrightarrow &y \in \bigcap_i CX_i(k) \nonumber \text{\hspace{48mm}}\blacksquare
\end{aligned}$ 
\begin{rem} \label{prop7}
The reverse inclusions need not hold in Lemmas \ref{prop5} and \ref{prop6}. 
Let $C=I$, $X_1=X_s$ and $X_2=X_{ns}$. 
$X_1 \cap X_2 = \emptyset$, but $X_1(k) \cap X_2(k)$ need not be empty. 
\footnote{Recall that the definition of the reachable set in $k$ steps assumes a nonempty initial set of states.}
\end{rem}
\begin{thm} \label{prop8}
If $X_{s_i}$ is $k-$ISO with respect to $X_{ns}$ for each $i$, then $\bigcap_i X_{s_i}$ is $k-$ISO with respect to $X_{ns}$.
\end{thm}

\emph{Proof}:
$\begin{aligned}[t]
&X_{s_i} \text{ }k-\text{ISO w.r.t. }X_{ns} \forall i \nonumber \\
\Leftrightarrow &CX_{s_i}(k) \subseteq CX_{ns}(k) \forall i \nonumber \\
\Rightarrow &\bigcap_i CX_{s_i}(k) \subseteq CX_{ns}(k) \nonumber \\
\Rightarrow &C(\bigcap_i X_{s_i}(k)) \subseteq CX_{ns}(k) \nonumber \\
\Leftrightarrow &\bigcap_i X_{s_i} \text{ is }k- \text{ISO w.r.t. }X_{ns} \nonumber \text{\hspace{30mm}}\blacksquare
\end{aligned}$ 
\begin{thm} \label{prop9}
If $X_{s}$ is $k-$ISO with respect to $X_{ns_i}$ for each $i$, then $CX_s(k) \subseteq \bigcap_i CX_{ns_i}(k)$. 
However, in general, $X_s$ is not $k-$ISO with respect to $\bigcap_i X_{ns_i}$.
\end{thm}
\begin{IEEEproof}
$\begin{aligned}[t]
&X_{s} \text{ }k-\text{ISO w.r.t. }X_{ns_i} \forall i \nonumber \\
\Leftrightarrow &CX_{s}(k) \subseteq CX_{ns_i}(k) \forall i \nonumber \\
\Rightarrow &CX_{s}(k) \subseteq \bigcap_i CX_{ns_i}(k) \nonumber 
\end{aligned}$ 

However, we can have $\bigcap_i X_{ns_i} = \emptyset$, which means $C(\bigcap_i X_{ns_i})(k)$ is undefined. 
\end{IEEEproof}

A similar result holds for weak opacity. 
\begin{thm} \label{prop10}
If $X_{s_i}$ is weakly $k-$ISO with respect to $X_{ns}$ for each $i$, then $\bigcup_i X_{s_i}$ is weakly $k-$ISO w.r.t. $X_{ns}$.
\end{thm}

\emph{Proof}:
$\begin{aligned}[t]
&X_{s_i} \text{ weakly }k-\text{ISO w.r.t. }X_{ns} \forall i \nonumber \\
\Leftrightarrow &CX_{s_i}(k) \bigcap CX_{ns}(k) \neq \emptyset \forall i \nonumber \\
\Rightarrow &\bigcup_i CX_{s_i}(k) \bigcap CX_{ns}(k) \neq \emptyset \nonumber \\
\Rightarrow &C(\bigcup_i X_{s_i}(k)) \bigcap CX_{ns}(k) \neq \emptyset \nonumber \\
\Leftrightarrow &\bigcup_i X_{s_i} \text{ is weakly }k- \text{ISO w.r.t. }X_{ns} \nonumber \text{\hspace{15mm}}\blacksquare
\end{aligned}$
\begin{rem} \label{prop11}
If $X_{s_i}$ is weakly $k-$ISO with respect to $X_{ns}$ for each $i$, then $\bigcap_i X_{s_i}$ need not be weakly $k-$ISO with respect to $X_{ns}$. 
That is, given $CX_{s_i}(k) \bigcap CX_{ns_i}(k) \neq \emptyset \forall i$, if $\bigcap_i X_{s_i} = \emptyset$, then $C(\bigcap_i X_{s_i})(k) \bigcap CX_{ns}(k)$ will not be defined.
\end{rem}
\section{Opacity and Reach-set Approximations} \label{ReachOpaApprox}
\subsection{Determining Reachable Sets of States}\label{ReachCompute}

The notions of opacity developed in this paper all rely on computing a reachable set of states at a time $k$. 
Exactly computing these sets for dynamical systems whose states take continuous values is usually impossible. 

A missing link in the works on computing approximations of reachable sets is that they do not study the effect of set operations (eg. intersection, subset containment) over these approximations on properties that rely on carrying out the same operations on the exact reachable sets. 
The notion of $k$-ISO proposed in this paper is an example of such a property (as shown in Theorem \ref{OpaIff}). 

For systems of the form in Equation (\ref{LTISys}), 
a major obstacle in exactly computing reachable sets of states is that of \emph{decidability} \cite{lafferriere2001symbolic}. 
Here, decidability refers to the existence of an efficient method to determine whether or not a given state, or a set of states, is reachable from a given set of states. 
The reachability problem was shown to be decidable for the special case of linear systems in which the $A$ matrix had a particular eigenstructure (constant and nilpotent, or diagonalizable with purely imaginary or rational real eigenvalues) in \cite{lafferriere2001symbolic}. 
However, the reachability problem for dynamical systems, in general, is undecidable. 

One means of exactly determining the set of reachable states at a time $k$ is from the Minkowski sum \footnote{The Minkowski sum of sets $S_1, S_2 \subseteq \mathbb{R}^d$ is $S_1 \oplus S_2 = \{s_1 + s_2 : ~s_1 \in S_1,~s_2 \in S_2\}$.} of the reachable states at time $k-1$ and the set of states at time $k$ (obtained from the system dynamics). 
However, the size of the representation grows at each step, and the problem becomes intractable for large time horizons.
This calls for the need to develop efficient algorithms that can compute `good' approximations of these reachable sets of states. 
There is a large body of work (\cite{girard2008efficient, girard2006efficient, kurzhanskiy2007ellipsoidal}) that has focused on this problem, and several techniques have been proposed to compute approximations of reachable states. 
The choice of technique is usually governed by how the initial set of states and inputs is specified. 
%

Furthermore, the approximations are tight in the sense that the approximate reachable set will touch the original reachable set at the points where inequalities defining the approximate sets (in terms of half-spaces or ellipsoids) attain equality. 
The reader is directed to Tables I and II in \cite{han2016enlarging} for a discussion on the merits of some of the methods to determine these sets.
\subsection{$k$-ISO and Reach-set Approximations}\label{ReachOpacity}

We now characterize $k$-ISO in terms of over- and under-approximations of sets of reachable sets at the time $k$. 
\begin{df}
Given a set $S$, $\overline{S}$ is an \emph{overapproximation} of $S$ if $S \subseteq \overline{S}$. 
$\underline{S}$ is an \emph{underapproximation} of $S$ if $\underline{S} \subseteq S$. 
\end{df}

$S$ is then the union over all its underapproximations, $\underline{S}$, and the intersection over all its overapproximations, $\overline{S}$. 
%
We further make the following assumption. 
\begin{ass}
Overapproximations and underapproximations of the reachable sets of states at time $k$, starting from $X_s$ and $X_{ns}$ at time $0$ have already been computed.
\end{ass}

Let $\overline{X}_s(k)$ and $\underline{X}_s(k)$ denote the overapproximation and underapproximation of the reachable set $X_s(k)$. 
Similarly, let $\overline{X}_{ns}(k)$ and $\underline{X}_{ns}(k)$ denote the overapproximation and underapproximation of the reachable set $X_{ns}(k)$. 
\begin{lm} \label{ReachLemma}
The following hold: 
\begin{enumerate}
\item $\underline{X}_s(k) \subseteq X_s(k) \subseteq \overline{X}_s(k)$;
\item $\underline{X}_{ns}(k) \subseteq X_{ns}(k) \subseteq \overline{X}_{ns}(k)$;
\item $C \underline{X}_s(k) \subseteq CX_s(k) \subseteq C \overline{X}_s(k)$;
\item $C \underline{X}_{ns}(k) \subseteq CX_{ns}(k) \subseteq C \overline{X}_{ns}(k)$.
\end{enumerate}
\end{lm}
\begin{IEEEproof}
The first two parts of the Lemma follow from the fact that the set $\underline{X}$ is an underapproximation of $X$, and $X$ is overapproximated by $\overline{X}$. 

The third and fourth parts can be concluded from a similar argument for the set $CX$. 
\end{IEEEproof}

Lemma \ref{ReachLemma} leads to a condition on the approximation of the reachable sets of states at time $k$, starting from $X_s$ and $X_{ns}$, if $X_s$ is $k$-ISO with respect to $X_{ns}$. 
\begin{pr} \label{kISOProp}
If $X_s$ is $k$-ISO with respect to $X_{ns}$, then $C\underline{X}_s(k) \subseteq C \overline{X}_{ns}(k)$.
\end{pr}
\begin{IEEEproof}
If $k$-ISO holds, then from Theorem \ref{OpaIff}) we have that $CX_s(k) \subseteq CX_{ns}(k)$. 
From parts $3)$ and $4)$ of Lemma \ref{ReachLemma}, we have that $C\underline{X}_s(k) \subseteq CX_s(k) \subseteq CX_{ns}(k) \subseteq C \overline{X}_{ns}(k)$. 
\end{IEEEproof}

\begin{figure}
 \centering
  \includegraphics[width=2.7 in]{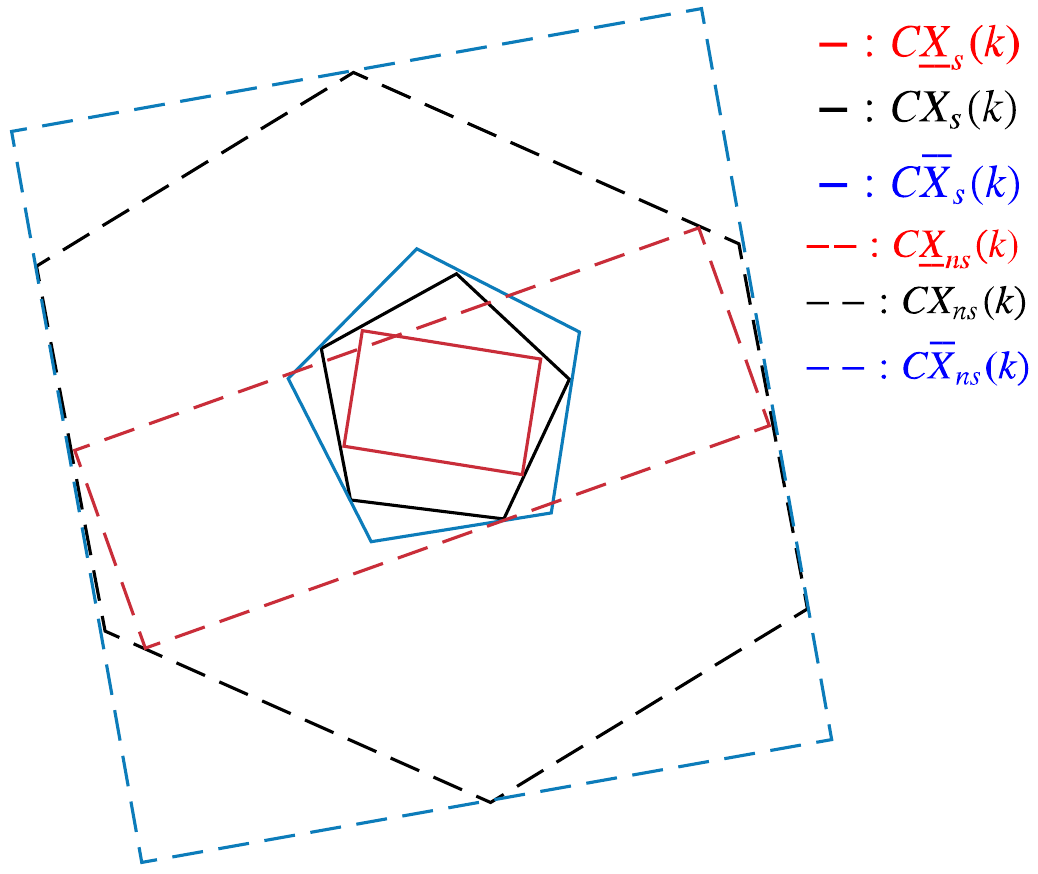} 
\caption{Representations of tight overapproximations and underapproximations of reachable sets for $k$-ISO. The figure illustrates Theorem \ref{OpaIff}, parts $3)$ and $4)$ of Lemma \ref{ReachLemma}, and Proposition \ref{kISOProp}. However, Proposition \ref{ReachProp} does not hold for this representation.} \label{kISOOverUnder} 
\end{figure}

Parts $3)$ and $4)$ of Lemma \ref{ReachLemma}, and Proposition \ref{kISOProp} are illustrated in Figure \ref{kISOOverUnder}. 

The next result is a converse of the previous argument, in that it presents a condition on the approximations of the reachable sets of states at time $k$ which will ensure that $k$-ISO holds. 
As one might conjecture, this condition will have to be somewhat stronger than the condition on the approximate reachable sets of states in Proposition \ref{kISOProp}. 
\begin{pr} \label{ReachProp}
If $C \overline{X}_s(k) \subseteq C \underline{X}_{ns}(k)$, then $X_s$ is $k$-ISO with respect to $X_{ns}$. 
\end{pr}
\begin{IEEEproof}
We have from Lemma \ref{ReachLemma} that $CX_s(k) \subseteq C \overline{X}_s(k) \subseteq C \underline{X}_{ns}(k) \subseteq CX_{ns}(k)$, which gives us $CX_s(k) \subseteq CX_{ns}(k)$. 
From Theorem \ref{OpaIff}, this is necessary and sufficient for $X_s$ to be $k$-ISO with respect to $X_{ns}$.
\end{IEEEproof}
\begin{rem}
At this point, it is important to realize the need to work with both over- and under-approximations of the reachable sets of states. 
The condition $C \overline{X}_s(k) \subseteq C \overline{X}_{ns}(k)$ does not reveal any useful information on whether $CX_s(k) \subseteq CX_{ns}(k)$, since this will depend on how tight the overapproximations are. 
This, however might not be enough to conclude whether $X_s$ is $k$-ISO w.r.t. $X_{ns}$ by using only the overapproximations of the sets of reachable states at time $k$, starting from $X_s$ and $X_{ns}$. 
\end{rem}

We now present a result on the computational complexity of verifying $k-$ISO in terms of the approximations of the reachable set of states. 
Assume that $L_{over}$ and $L_{under}$, respectively, denote the number of overapproximations and underapproximations of $X_s(k)$ and $X_{ns}(k)$. 
\begin{pr}\label{CompComp}
Let $L:=L_{over}+L_{under}$. 
Then, the computational complexity of verifying that $X_s$ is $k-$ISO with respect to $X_{ns}$ is $c_1 k L n^3 + c_2 p n+c_3$, for constants $c_1, c_2, c_3$.
\end{pr}
\begin{IEEEproof}
The term $c_1 k L n^3$ is an upper bound on computing the reachable sets at time $k$. 
These will depend on the complexity of computing $A^k$, which is polynomial in the dimension of the state space, $n$, and is typically of the order of $n^3$. 
This is done a total of $L$ times for time $k$. 
The second term is the computational complexity of the matrix-vector multiplication $Cx$, where $C \in \mathbb{R}^{p \times n}$ and $x \in \mathbb{R}^n$. 
This term does not depend on $L$ since we assume that this operation is performed on the union of the underapproximations, and intersection of the overapproximations, which we assume to take a constant amount of time. 
The third term represents other constant factors that result from performing these operations.
\end{IEEEproof}
\begin{rem}
In Proposition \ref{CompComp}, we refrain from using the more standard notation, $O(\cdot)$, for computational complexity, since $k, L,n$ may not be of comparable magnitudes.
\end{rem}

The next result presents a necessary and sufficient condition for $X_s$ to be $k$-ISO with respect to $X_{ns}$ in terms of the exact computations of the reachable sets of states at time $k$, starting from $X_s$ and $X_{ns}$. 

Before presenting the result, we require some definitions. 
From Equation (\ref{LTISys}), observe that $C: \mathbb{R}^n \rightarrow \mathbb{R}^p$ is a linear map from the set of states to the set of outputs. 
\begin{df}
The \emph{inverse image} of $C$ with respect to a set $\mathcal{Y} \subseteq \mathbb{R}^p$, denoted $C^{-1}$, is: 
\begin{align*}
C^{-1}\mathcal{Y}&:=\{x | x \in \mathbb{R}^n, Cx \in \mathcal{Y}\}
\end{align*}
\end{df}
\begin{df}
Consider a reachable set of states at time $k$, denoted $X(k)$. 
Then, the \emph{backward reachable set} at time $0$ starting from $X(k)$, denoted $Pre_0(X(k))$, is the set of states at time $0$ from which one can reach a state in $X(k)$ at time $k$ by applying a set of $k$ control inputs:
\begin{align*}
Pre_0(X(k)):=&\{x \in X_0| \exists u(0),\dots,u(k-1) \text{ s.t. } x(i+1) = Ax(i)\\&+Bu(i), i=0,\dots,k-1; x(0) = x; x(k) \in X(k)\}
\end{align*}
\end{df}

The next result characterizes $k$-ISO in terms of the inverse image of $C$, and the backward reachable set at time $0$, starting from reachable sets of states at time $k$. 
\begin{thm}\label{ReachOpaThm}
$X_s$ is $k$-ISO with respect to $X_{ns}$ if and only if the following two conditions are satisfied: 
\begin{enumerate}
\item $(Pre_0[C^{-1}(CX_s(k))]) \cap X_{ns} \neq \emptyset$;
\item $X_s \subset Pre_0[C^{-1}(CX_{ns}(k))]$.
\end{enumerate}
\end{thm}
\begin{IEEEproof}
First, assume that Conditions $1)$ and $2)$ of Theorem \ref{ReachOpaThm} hold. 
Note that the result of $C^{-1}(CX_s(k))$ is a subset of $\mathbb{R}^n$ given by $X_s(k) \cup Z(k)$, where $Z(k): = \{z|z \notin X_s(k), Cz \in CX_s(k)\}$. 
Let $X'(k): =  X_s(k) \cup Z(k)$. 
Then, $Pre_0[X'(k)]$ will give the set of states at time $0$ from which $X'(k)$ can be reached at time $k$. 
Condition $1)$ being satisfied means that there exists a nonsecret initial state in $X_{ns}$ from which by applying some set of $k$ control inputs, we can reach a state at time $k$ such that $Cx \in CX_s(k)$. 
Additionally, if Condition $2)$ is also satisfied, then for every initial secret state, and for every set of $k$ control inputs starting from $X_s$, there is a nonsecret initial state and a set of $k$ controls such that the outputs at time $k$ are indistinguishable. 
This, by Definition \ref{OpacityDefn}, is $k$-ISO. 

Now, assume that $X_s$ is $k$-ISO with respect to $X_{ns}$. 
Then, from Theorem \ref{OpaIff}, we have $CX_s(k) \subseteq CX_{ns}(k)$. 
Therefore, there exists a state in $X_{ns}$ from which we can reach a state in $X_{ns}(k)$ by applying $k$ control inputs, such that $Cx \in CX_{s}(k)$ for some $x \in X_{ns}(k)$. 
This is condition $1)$. 
Condition $2)$ follows from the fact that for every state in $X_s$, by applying any set of $k$ controls, it must be the case that $Cx \in CX_{ns}(k)$, where $x \in X_s(k)$. 
\end{IEEEproof}

Although Theorem \ref{ReachOpaThm} presents conditions to establish $k$-ISO in terms of the exact computations of the reachable sets of states, overapproximations and underapproximations of these sets can be used as a means to enforce opacity on a secret specification that is not $k$-ISO w.r.t. a given nonsecret specification. 
This can be achieved by `pruning' the initial secret specification to remove states that might be `vulnerable' to revealing the secret (and hence will not be opaque), in order to get a new specification of the secret (which will be a subset of the original specification) that will be $k$-ISO with respect to the nonsecret specification, $X_{ns}$. 
As an example, assume that $CX_s(k) \not\subset CX_{ns}(k)$, but $CX_s(k) \cap CX_{ns}(k) \neq \emptyset$. 
This means that there exist initial secret states in $X_s$ and/ or controls starting from these states that will reveal the secret at time $k$.

We now present a way to prune the initial secret specification that will ensure opacity of a secret specification that will be a subset of the original specification. 
\begin{pr}
Let $CX_s(k) \not\subset CX_{ns}(k)$, and $CX_s(k) \cap CX_{ns}(k) \neq \emptyset$. 
Define a new secret specification to be $X_s' := X_s \cap Pre_0[CX_s(k) \cap CX_{ns}(k) ]$. 
Then $X_s'$ will be $k$-ISO with respect to $X_{ns}$. 
\end{pr}
\begin{IEEEproof}
Notice that 
$Pre_0[CX_s(k) \cap CX_{ns}(k)]$ eliminates those initial states from which a state $x$ can be reached at time $k$ such that $Cx \in (CX_s(k) \setminus CX_{ns}(k))$. 
The new secret specification is obtained by considering only those states in $X_s$ that will lead to a state $x$ such that $Cx \in CX_{ns}(k)$ (thus satisfying Condition $2$ of Theorem \ref{ReachOpaThm}). 
\end{IEEEproof}

A \emph{good} candidate initial secret set of states will not require too much `pruning' in order to ensure that it is $k$-ISO w.r.t. a nonsecret set of initial states. 
That is, this set will not contain states that are vulnerable to being revealed to an adversary. 
A set $X_s$ will be a \emph{good} candidate initial secret set if $C \underline{X}_s(k) \subseteq C \underline{X}_{ns}(k)$ and $C \overline{X}_s(k) \cap C \underline{X}_{ns}(k) \neq \emptyset$. 
The \emph{best} candidate initial secret set will require no pruning in order to ensure opacity. 
$X_s$ will be the \emph{best} candidate initial secret set if $C \overline{X}_s(k) \subseteq C \underline{X}_{ns}(k)$.
\begin{rem}
A \emph{good} (respectively, the \emph{best}) candidate set mentioned above is `good' (the `best') in the literal sense of the meaning of the word; we are yet to formulate a mathematically precise notion of this set. 
\end{rem}
%
%
%
\section{Opacity and Output Controllability} \label{OpacityOutputCont}

A state of the system is said to be controllable if we can find an input that transfers the state to the origin in finite time. 
While there are several interesting results in the literature that relate controllability of a dynamical system to other properties of interest, the notion of \emph{output controllability} has been largely overlooked. 
Output controllability is the ability to transfer the state of the system such that the output corresponding to the state at some finite time is zero. 
It is easy to see that while controllability implies output controllability, the reverse argument need not necessarily hold. 
This section establishes an equivalence between $k$-ISO and output controllability. 
%
\begin{df}
A state $x$ of (\ref{LTISys}) is \emph{controllable} on
$[0,k_f]$ if there exists a control
sequence $\{u(\cdot)\}$ that transfers the state of the system from $x(0)=x$ to $x(k_f)=0$. 
\end{df}

%
The output of (\ref{LTISys}) at time $k$ is given by:
\begin{align}
y(k)&=CA^kx(0)+\sum_{j=0}^{k-1}CA^{k-j-1}Bu(j) \nonumber
\end{align} 
\begin{df}
%
A state $x$ of (\ref{LTISys}) is \emph{output controllable} on $[0,k_f]$ if there
exists a control sequence $\{u(\cdot)\}$ that transfers the system
from $x(0)=x$ to $x(k_f)$, so that $y(k_f)=0$.
\end{df}
The main result of this section indicates that $X_s$ being $k$-ISO with respect to $X_{ns}$ ensures that there exists a state that is output controllable. 
We then prove that the converse holds, under an additional assumption.
\begin{thm} \label{OutContI}
Let $X_s$ be $k-$ISO with respect to $X_{ns}$. Then, there exists a state of (\ref{LTISys}) that is output controllable on $[0,k]$. 
Further, if $k-$ISO is established for the pair $(x_s(0), x_{ns}(0)) \in X_s \times X_{ns}$ (and appropriate control sequences $\{u_s(\cdot)\}$ and $\{u_{ns}(\cdot)\}$), then the control sequence $u(i)=u_s(i)-u_{ns}(i)$, $i=0,1,\dots,k-1$, will achieve output controllability for the initial state $x(0)=x_s(0)-x_{ns}(0)$.
\end{thm}
\begin{IEEEproof} $k-$ISO implies $y_s(k)=y_{ns}(k)$
for appropriate $x_s(0), \{u_s(\cdot)\},x_{ns}(0)$ and
$\{u_{ns}(\cdot)\}$. 
Setting $x(0)=x_s(0)-x_{ns}(0)$ and $u(i)=u_s(i)-u_{ns}(i)$, $i=0,1,\dots,k-1$ in the dynamics of (\ref{LTISys})  ensures $y(k)=0$, thus achieving output
controllability of the state $x(0)=x_s(0)-x_{ns}(0)$. 
\end{IEEEproof}
\begin{thm} \label{OutContII}
Let (\ref{LTISys}) be output controllable in $k$ steps for a set of states $X_{oc}(0) \setminus \{0\}$ and controls $\{U(\cdot)\}$. 
Let $X_1$ and $X_2$ be sets such that every $x_1 \in X_1$ can be written as $x+x_2$, where $x \in X_{oc}(0) \setminus \{0\}$ and $x_2 \in X_2$. 
Then, $X_1$ is strongly $k-$ISO with respect to $X_2$.
\end{thm}
\begin{IEEEproof} Output controllability ensures that:
\begin{align}
y(k)=CA^kx(0)+\sum_{j=0}^{k-1}CA^{k-j-1}BU(j)=0 \label{OutCont}
\end{align} 
For any control sequence $\{u_1(\cdot)\}$, the output at time $k$, starting from any $x_1(0) \in X_1$ is: 
\begin{align}
y_1(k)&=CA^kx_1(0)+\sum_{j=0}^{k-1}CA^{k-j-1}Bu_1(j)  \nonumber
\end{align}
The output at time $k$ starting from $x_2(0) \in X_2$ with the control sequence $\{u_1(\cdot)-U(\cdot)\}$ is: 
\begin{align}
y_2(k)&=CA^kx_2(0)+\sum_{j=0}^{k-1}CA^{k-j-1}B[u_1(j)-U(j)] \nonumber
\end{align}
Every $x_1 \in X_1$ can be written as $x+x_2$, where $x \in X_{oc}(0) \setminus \{0\}$, $x_2 \in X_2$. This and Equation (\ref{OutCont}) gives $y_1(k)=y_2(k)$. 

Thus, for any $x_1 \in X_1$ and any control sequence starting from $x_1$, there exist $x_2 \in X_2$ and another control sequence such that the outputs after $k$ steps are the same. 
This is strong $k-$ISO with $X_s=X_1$ and $X_{ns}=X_2$. 
\end{IEEEproof}
\section{Decentralized Opacity} \label{OpacityDecent}

In this section, we define several notions of decentralized opacity in the presence of multiple adversaries. 
The presence or absence of collusion among the adversaries, and the presence or absence of a coordinator that aggregates information based on the adversaries' observations, is the distinguishing feature, and a definition of decentralized opacity is proposed in each case. 
As in the single adversary case, every adversary is assumed to have knowledge of the initial sets of secret and nonsecret states, $X_s$ and $X_{ns}$, the system model $(A,B)$, and its own observation map $C_i$, and is assumed to have unlimited computing power. 
The system model is described in Equation (\ref{LTISysMult}):
\begin{align}
x(t+1)&=Ax(t)+Bu(t) \nonumber \\
x(0)&=x_0 \in X_0 \nonumber \\
y_i(t)&=C_ix(t); \text{  }i=1,2,\dots,l \label{LTISysMult}
\end{align}  
where $x \in \mathbb{R}^n, u \in \mathbb{R}^m, y_i \in \mathbb{R}^{p_i}$, and $A,B,C_i$ are matrices of appropriate dimensions containing real entries. 
In the sequel, we will assume that all of the adversaries observe the system at the same time instants in the set $\mathcal{K}$. 

\begin{rem}
The scenarios that we study when there is more than one adversary are not enforced on the adversaries. 
Instead, we seek to account for the different ways in which these adversaries could act. 
This is a subtle, yet important distinction. 
The definition of decentralized opacity which will apply will be according to the distinguishing feature described at the start of this section. 
\end{rem}
\subsection{No Coordinator, No Collusion} \label{nn}

The agents are assumed to not communicate with each other, and there is no centralized coordinator. 
Opacity of the secret will be achieved when it is simultaneously opaque with respect to every adversary. 
\begin{df} 
For system (\ref{LTISysMult}), given $X_s, X_{ns} \subseteq X_0$ and $k \in \mathcal{K}$, $X_s$ is
  \emph{strongly decentralized $k-$ISO} with respect to $X_{ns}$ if for all $x_s(0) \in X_s$ and for every sequence of admissible controls $u_s(0), \dots , u_s(k-1)$, there exist an $x_{ns}(0) \in X_{ns}$, and a sequence of admissible controls
  $u_{ns}(0), \dots, u_{ns}(k-1)$ such that ${y_s}_i(k)={y_{ns}}_i(k)$ for all $i \in \{1,2,\dots,l\}$. 
  
$X_s$ is \emph{strongly decentralized $\ca{K}-$ISO} with respect to $X_{ns}$ if it is strongly decentralized $k-$ISO for all $k \in \ca{K}$.
\end{df}
As in the single adversary case, we have a necessary and sufficient condition for decentralized opacity in terms of sets of reachable states in $k$ steps. 
\begin{thm} \label{DecentOpaIff}
The following hold: 
\begin{enumerate}
\item $X_s$ is strongly decentralized $k-$ISO w.r.t. $X_{ns}$ if and only if $C_iX_s(k) \subseteq
C_iX_{ns}(k)$ for all $i \in \{1,2,\dots,l\}$. 
\item $X_s$ is strongly decentralized $\ca{K}-$ISO w.r.t. $X_{ns}$ if and only if $C_iX_s(k) \subseteq
C_iX_{ns}(k)$ for all $k \in \ca{K}$, and for all $i \in \{1,2,\dots,l\}$.
\end{enumerate}
\end{thm}
\begin{IEEEproof}
The proof follows from Theorem \ref{OpaIff}. 
\end{IEEEproof}

The following result explores the relationship between decentralized $k-$ISO for a set of adversaries and $k-$ISO for a single adversary with an aggregated observation map. 
\begin{pr} \label{AggOpa}
$X_s$ is strongly decentralized $k-$ISO with respect to $X_{ns}$ and adversaries with observation maps $C_1,\dots,C_l$ if $X_s$ is strongly $k-$ISO with respect to $X_{ns}$ for the single adversary with the aggregated observation map $\bar{C}:=\begin{pmatrix} C_1^T & C_2^T & \dots & C_l^T \end{pmatrix}^T$. 
\end{pr}
\begin{IEEEproof} 
$X_s$ strongly $k-$ISO with respect to $X_{ns}$ is equivalent to $\bar{C}X_s(k) \subseteq \bar{C}X_{ns}(k)$. 
This means that for every $x_s(k) \in X_s(k)$, there exists an $x_{ns}(k) \in X_{ns}(k)$ such that $C_1x_s(k)=C_1x_{ns}(k)$, $\dots$, $C_lx_s(k)=C_lx_{ns}(k)$. 
Thus, we have $C_iX_s(k) \subseteq C_iX_{ns}(k)$ for all $i \in \{1,\dots,l\}$, which is equivalent to $X_s$ being strongly decentralized $k-$ISO w.r.t. $X_{ns}$. 
\end{IEEEproof}

It is to be noted that strong decentralized $k-$ISO need not necessarily ensure strong $k-$ISO with respect to an adversary with the aggregated observation map since, the nonsecret states in $X_{ns}(k)$  and the corresponding control sequence for each adversary may be different. 
\subsection{With Coordinator, No Collusion} \label{yn}

Here, we assume that there is a coordinator, whose role is to poll the observations of each adversary, and decide on \emph{co-opacity} according to some (predefined) rule. 
The coordinator does not have knowledge of the system model or the adversaries' observation maps. 
In fact, our model is such that the coordinator cannot do any better even if it knows the system model or the observation maps. 
It can be viewed as an agent whose role is to ensure that the whole is greater than the sum of its parts. 

Formally, the coordinator communicates to the adversaries the time instants $\ca{K}$, at which the system needs to be observed. 
At each $k \in \ca{K}$, agent $i$ observes $y_i(k)=C_ix(k)$. 
The agents communicate $\phi_i(y_i(k))$ to the coordinator, where $\phi_i: \bb{R}^{p_i} \rightarrow 2^{\bb{R}^{n}\times \bb{R}^n}$ is defined as:
\begin{align}
\phi_i(y_i(k)):=\{(x^1,x^2) \in X_s(k) \times X_{ns}(k) : C_ix^1=C_ix^2=y_i(k)\} \nonumber
\end{align}

Thus, $\phi_i(\cdot)$ returns secret-nonsecret state pairs that give the same output $y_i(k)$ at time $k$. 

The coordinator then computes a function $\Psi (k):=\Psi (\phi_1(y_1(k)), \dots, \phi_l (y_l(k)))$, where $\Psi : (2^{\bb{R}^{n}\times \bb{R}^n})^l \rightarrow 2^{\bb{R}^{n}\times \bb{R}^n}$. 
Thus, the coordinator plays the role of gathering the outputs of the observations of each adversary, and composing them to then decide on opacity. 
An example of a valid coordinator function is $\Psi (k)= \bigcup_i (\phi_i(C_ix(k)))$. 

\begin{figure}
 \centering
  \includegraphics[width=2.7 in]{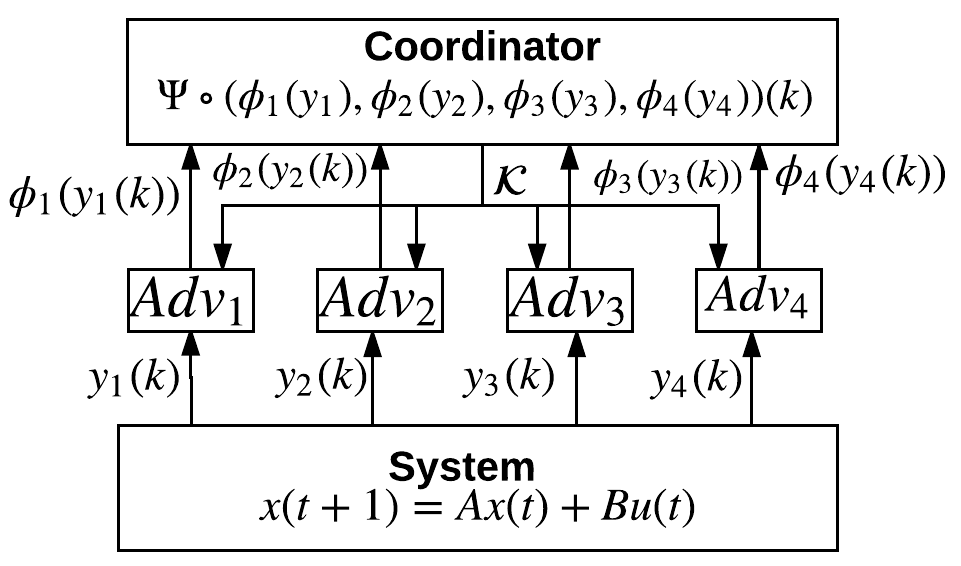} 
\caption{Coordinated Decentralized Opacity} \label{CoISO} 
\end{figure}

The scheme is shown in Figure \ref{CoISO} for four adversaries. 
\begin{df} 
For system (\ref{LTISysMult}), given $X_s, X_{ns} \subseteq X_0$ and $k \in \mathcal{K}$, $X_s$ is
  \emph{strongly co-$k-$ISO} with respect to $X_{ns}$ and $\Psi$ if for all $x_s(0) \in X_s$ and for every sequence of admissible controls $u_s(0), \dots , u_s(k-1)$, there exist an $x_{ns}(0) \in X_{ns}$, and a sequence of admissible controls
  $u_{ns}(0), \dots, u_{ns}(k-1)$ such that $\Psi(k)$ is nonempty. 
  
$X_s$ is \emph{strongly co-$\ca{K}-$ISO} with respect to $X_{ns}$ and $\Psi$ if it is strongly co-$k-$ISO for all $k \in \ca{K}$.
\end{df} 

Before presenting the main result of this section, we provide an alternative characterization of strong $k-$ISO in terms of the map $\phi$ (the subscript on $\phi_i$ is dropped since we consider only a single adversary in this case). 
Further, it is important to note that the functions $\phi_i$ and $\Psi$ return a set of pairs of states at time $k$. 
This information needs to be used to determine opacity of the initial set of secret states with respect to the initial set of nonsecret states. 

We extend the definition of $\phi$ to sets of outputs at time $k$. 
Let $\phi(CX(k)):=\bigcup \{\phi(y(k)): [y(k)=Cx(k)] \wedge [x(k) \in X(k)]\}$. 
For $(x^1_i,x^2_j) \in X_s(k) \times X_{ns}(k)$, in a slight abuse of notation, we treat each $x^1_i$ and $x^2_j$ as a set. 
This will allow us to define $\bigcup_{i,j} (x^1_i,x^2_j):=(\bigcup_i x^1_i, \bigcup_j x^2_j)$, where $\bigcup_i x^1_i \subseteq X_s(k)$, and $\bigcup_j x^2_j \subseteq X_{ns}(k)$. 
\begin{pr}
$X_s$ is strongly $k-$ISO with respect to $X_{ns}$ if and only if $\phi(CX_s(k))=(X_s(k),X'_{ns}(k))$, where $X'_{ns}(k):=\{x \in X_{ns}(k) : Cx \in CX_{s}(k)\}$.
\end{pr}
\begin{IEEEproof}
Let strong $k-$ISO hold. 
Then, $CX_s(k) \subseteq CX_{ns}(k)$ (Theorem \ref{OpaIff}), and $\phi(CX_s(k))=(X_s(k),X'_{ns}(k))$, where $X'_{ns}(k)$ is as defined above. 

If $\phi(CX_s(k))=(X_s(k),X'_{ns}(k))$, then $\forall x^1 \in X_s(k)$, $\exists x^2 \in X'_{ns}(k) \subseteq X_{ns}(k)$ such that $Cx^1(k)=Cx^2(k)$. 
This gives $CX_s(k) \subseteq CX_{ns}(k)$, and hence strong $k-$ISO.
\end{IEEEproof}

The above result says that strong $k-$ISO holds if and only if the first component of $\phi(\cdot)$ when acting on the set of secret outputs at time $k$ is the entire set of reachable states at time $k$, starting from $X_s$. 
Further, it also determines the states in $X_{ns}(k)$ that ensure strong $k-$ISO. 
\begin{thm} \label{CoOpaIff}
$X_s$ is strongly co-$k-$ISO with respect to $X_{ns}$ and $\Psi$ if and only if $\Psi (\phi_1(C_1X_s(k)), \dots, \phi_l (C_lX_s(k)))=(X_s(k),X'_{ns}(k))$, where $X'_{ns}(k) \subseteq X_{ns}(k)$. 
\end{thm}
\begin{IEEEproof}
The proof follows from the previous result, and the definition of co$-k-$ISO. 
The major difference is that in this case, the first component of $\phi_i(C_iX_s(k))$ can be a subset of $X_s(k)$. 
However, the coordinator function $\Psi$ must be such that its first component is $X_s(k)$. 
\end{IEEEproof}

Thus, $X_s$ can be strongly co$-k-$ISO w.r.t. $X_{ns}$ though strong $k-$ISO might not hold for any single adversary. 
\subsection{No Coordinator, With Collusion} \label{ny}

\begin{figure}
 \centering
  \includegraphics[width=2.7 in]{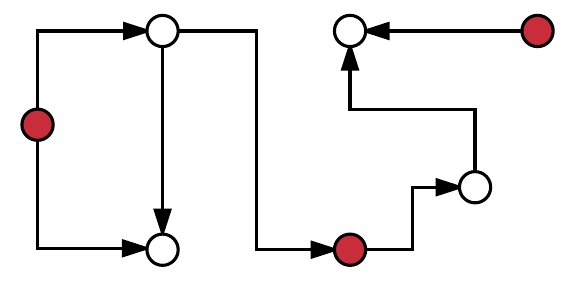} 
\caption{Vertices in red form a directed dominating set} \label{DirDom} 
\end{figure}

%
In this case, there is no coordinator, but the adversaries are assumed to communicate among themselves. 
This is a new approach, and has not been studied for DESs. 
The communication structure is represented by a directed graph $\ca{G}$, whose vertices are the adversaries, and $\ca{G}$ has an edge directed from $i$ to $j$ if adversary $j$ can receive information from adversary $i$. $j$ is then called a \emph{neighbor} of $i$. 
The goal of the adversaries is to ensure, using the coordination structure, that $X_s$ is not $k-$ISO with respect to $X_{ns}$ for each of them. 
To this end, we introduce the following definitions:
\begin{df}
For the system (\ref{LTISysMult}), given $X_s, X_{ns} \subseteq X_0$ and $k \in \mathcal{K}$, $X_s$ is \emph{strongly not $k-$ISO} with respect to $X_{ns}$ if $X_s$ is not strongly $k-$ISO w.r.t. $X_{ns}$ for every adversary. 
\end{df}
\begin{df}
Given a graph $\ca{G}=(\ca{V},\ca{E})$, with vertices $\ca{V}$ and edges $\ca{E} \subset \ca{V} \times \ca{V}$, $D \subset \ca{V}$ is a \emph{dominating set} if every vertex not in $D$ has a neighbor in $D$. 

Given a directed graph $\ca{G}=(\ca{V},\ca{E})$, $D \subset \ca{V}$ is a \emph{directed dominating set}(red vertices in Figure \ref{DirDom}) if every vertex not in $D$ has an incoming edge from some vertex in $D$, that is, $[\forall u \in \ca{V} \setminus D$, $\exists v \in D$ such that $(v \rightarrow u) \in \ca{E}]$.  
\end{df}

At each $k \in \ca{K}$, each adversary observes $y(k)$, determines if $k-$ISO holds or not, and communicates $(C_i, <k-$ISO status$>_i)$ to its neighbors in $\ca{G}$. 
If $<k-$ISO status$>_i=0$, i.e. $k-$ISO does not hold for adversary $i$, then a neighbor $j$ of $i$ in $\ca{G}$ adopts $C_i$ as its observation map if $<k-$ISO status$>_j \neq 0$. 
This scheme can be interpreted as a dynamic version of $k-$ISO, in which the adversaries change their observation maps at times $k \in \ca{K}$ depending on the $k-$ISO status of their neighbors in $\ca{G}$. 
\begin{ass}
The time required for the adversaries to communicate amongst themselves is much less than the time scale of the system. 
\end{ass}
We present a sufficient condition to achieve strong non-opacity without requiring non-opacity w.r.t. every adversary using the communication scheme described above. 
\begin{thm}\label{ThmDirDom}
For the system (\ref{LTISysMult}), $X_s$ is strongly not $k-$ISO with respect to $X_{ns}$ if the set of adversaries for which $X_s$ is not strongly $k-$ISO with respect to $X_{ns}$ is a directed dominating set of $\ca{G}$. 
\end{thm}
\begin{IEEEproof}
Each adversary communicates $(C_i, <k-$ISO status$>_i)$ to its neighbors in $\ca{G}$. 
Thus, if $k-$ISO does not hold for some adversary $i$, then its neighbors will also adopt the same $C_i$ matrix at time $k$. 
The result follows from the definition of a directed dominating set. 
\end{IEEEproof}
%
%
%
%
%
%
\section{$\epsilon$-Opacity} \label{OpacityEpsilon}

The condition that the output at times $k$ starting from every state in $X_s$ be equal to that got by starting from a state in $X_{ns}$ is quite strong. 
In this section, we postulate that (a form of) opacity (for the single adversary case) will still hold if the outputs differ by an arbitrary amount. 

\begin{df}
  For system (\ref{LTISys}), given $X_s, X_{ns} \subseteq X_0$, $k \in \mathcal{K}$, and $\epsilon \geq 0$, $X_s$ is
  \emph{strongly $\epsilon- k-$ISO} with respect to $X_{ns}$ if for all $x_s(0) \in X_s$ and for every sequence of admissible controls $u_s(0), \dots , u_s(k-1)$, there exist an $x_{ns}(0) \in X_{ns}$, and a sequence of admissible controls
  $u_{ns}(0), \dots, u_{ns}(k-1)$ such that $\lVert y_s(k)-y_{ns}(k) \rVert_2 \leq \epsilon$. 
  
$X_s$ is \emph{strongly $\epsilon -\ca{K}-$ISO} with respect to $X_{ns}$ if it is strongly $\epsilon -k-$ISO for all $k \in \ca{K}$.
\end{df}
%

\begin{figure}
 \centering
  \includegraphics[width=2.3 in]{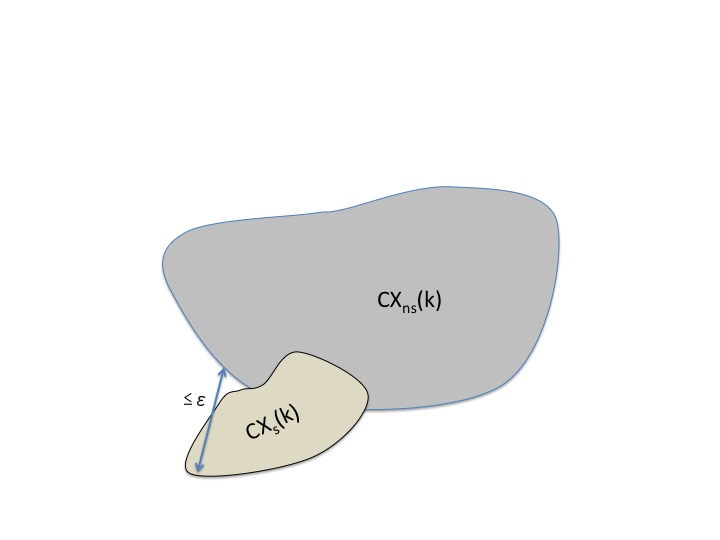} 
\caption{Representation of $\epsilon-k-$ISO} \label{epsISO} 
\end{figure}

Let $z$ be a point, and $S$ be a set. 
Then, the distance of $z$ from $S$ is defined as $dist(z,S):=inf\{dist(z,s) | s \in S\}$. 
\begin{thm} \label{EpsOpaIff}
The following hold: 
\begin{enumerate}
\item $X_s$ is strongly $\epsilon-k-$ISO w.r.t. $X_{ns}$ if and only if :
\begin{align}
\max_{z \in CX_s(k)} dist(z,CX_{ns}(k)) \leq \epsilon \label{EpsOpa}
\end{align}
That is, the farthest a point in $CX_s(k)$ can be from $CX_{ns}(k)$ is $\epsilon$. 
This is shown in Figure \ref{epsISO}.
\item $X_s$ is strongly $\epsilon-\ca{K}-$ISO with respect to $X_{ns}$ if and only if (\ref{EpsOpa}) holds for all $k \in \ca{K}$.
\end{enumerate}
\end{thm}
\begin{IEEEproof}
The proof of this result follows from the definition of $\epsilon-k$-ISO and Theorem \ref{OpaIff}. 
\end{IEEEproof}

Notice that $\epsilon=0$ in Theorem \ref{epsISO} corresponds to the definition of strong $k$-ISO. 
This notion of $\epsilon-k$-ISO can be used to quantify how opaque a secret specification is. 
Differential privacy was proposed in \cite{dwork2008differential, dwork2011differential} to provide probabilistic guarantees on the indistinguishability of datasets. 
It remains to be seen if a connection can be made between this and $\epsilon-k$-ISO. 
An important distinction between the frameworks is that ours is deterministic, while that considered in differential privacy is stochastic. 
\section{Opacity for Nonlinear Systems} \label{OpacityNonLin}

Thus far, we have analyzed opacity in the context of linear systems. 
In this section, we extend the analysis to discrete-time nonlinear systems (DTNLSs), and formulate a notion of opacity for such systems. 
Consider the DTNLS: 
\begin{align}
x(t+1)&=f(x(t),u(t)) \nonumber \\
x(0)&=x_0 \in X_0 \nonumber \\
y(t)&=h(x(t)) \label{NonLinSys}
\end{align}
where $x \in \mathbb{R}^n, u \in \mathbb{R}^m, y \in \mathbb{R}^p$, and $f(\cdot, \cdot)$ and $h(\cdot)$ are sufficiently smooth functions with $h(0_n)=0_p$, where $0_*$ is the $1 \times *$ vector of zeroes. 

Before studying opacity for systems of this kind, we take a brief detour to make note of some of the relevant literature. 
In a series of papers, the authors of \cite{lee1986approximate}, \cite{lee1987input}, and \cite{lee1987linearization} derived conditions under which a DTNLS could be equivalently represented as a discrete-time linear system. 
A geometric analysis of controllability of DTNLSs in terms of Lie algebras of vector fields was presented in \cite{jakubczyk1990controllability}. 
A linear algebraic framework for the analysis of synthesis problems in DTNLSs was proposed in \cite{grizzle1993linear}, where the notion of the rank of an analytic discrete-time system was developed. 
In a more recent work, the authors of \cite{jiang2001input} studied input-to-state stability properties of DTNLSs, using well established notions of input-to-state stability from the continuous time version. 
\begin{rem}  \cite{grizzle1993linear} 
The analysis of continuous time nonlinear systems is largely focused on systems that are affine in the input (that is, of the form $\dot{x}(t)=f(x(t))+g(x(t))u(t)$). 
The advantages that such a model offers is twofold: i) the derivatives of the output depend polynomially on the inputs and their derivatives, and ii) the vector fields involved have a nice structure (a drift term and $m$ control vector fields). 
Moreover, this class of systems is general enough to model many practical nonlinear systems. 
However, the class of discrete-time nonlinear systems can also potentially include versions of continuous time systems that are sampled in time, which necessitates considering the more general form of the DTNLS in Equation (\ref{NonLinSys}). 
\end{rem}

In this section, we extend our formulation of $k$-ISO for DT-LTI systems to DTNLSs. 
The assumptions are similar to those in Section \ref{OpacityLTI}. 
%
We have the following definition:
\begin{df} \label{OpacityDefnNonlin}
For the system (\ref{NonLinSys}), given $X_s, X_{ns} \subset X_0$ and $k \in \ca{K}$, $X_s$ is \emph{strongly $k$-initial state opaque ($k$-ISO)} with respect to $X_{ns}$ if for every $x_s(0) \in X_s$ and for every sequence of admissible controls $u_s(0),\dots,u_s(k-1)$, there exist an $x_{ns}(0) \in X_{ns}$ and a sequence of admissible controls $u_{ns}(0),\dots,u_{ns}(k-1)$ such that $y_s(k)=y_{ns}(k)$. 

$X_s$ is \emph{strongly $\ca{K}$-ISO} with respect to $X_{ns}$ if $X_s$ is strongly $k$-ISO with respect to $X_{ns}$ for all $k \in \ca{K}$. 
\end{df}

This definition of $k$-ISO is weaker than the notion of \emph{indistinguishability} in the nonlinear control literature (see for example, \cite{hermann1977nonlinear}), where two states are said to be indistinguishable if the outputs are identical at each instant of time for every admissible control sequence. 

In this case also, since only the set of possible inputs that can be applied at each time step in $[0,k]$ is available, we can establish conditions for $k$-ISO in terms of sets of reachable states. 
Let $U_{s}^k:=\{u_{s}(0),\dots,u_{s}(k-1)\}$ and $U_{ns}^k:=\{u_{ns}(0),\dots,u_{ns}(k-1)\}$. 
Let $X_s(k)$ and $X_{ns}(k)$ denote the sets of states reachable in $k$ steps, starting at time $0$ from \emph{nonempty} sets $X_s$ and $X_{ns}$ respectively. 
That is, 
\begin{align}
X_s(k) &= \bigcup _{x_0 \in X_s} \bigcup _{U_{s}^k}\{x: x(i+1)=f(x(i),u(i)), \forall i <k \} \label{secreach} \\
X_{ns}(k)&= \bigcup _{x_0 \in X_{ns}} \bigcup_{U_{ns}^k}\{x: x(i+1)=f(x(i),u(i)), \forall i <k \} \label{nonsecreach}
\end{align}

The set of outputs at time $k$ can be represented by
\begin{align}
h(X(k))&:=\{y: y = h(x), x \in X(k)\}
\end{align}

We have the following result:
\begin{thm}
For the system (\ref{NonLinSys}), $X_s$ is strongly $k$-ISO with respect to $X_{ns}$ if and only if $h(X_s(k)) \subseteq h(X_{ns}(k))$.  
\end{thm}
\begin{IEEEproof}
The proof of this result can be developed similar to that of Theorem \ref{OpaIff}, and is omitted for brevity. 
\end{IEEEproof}

We have shown that there is a connection between $k$-ISO and output controllability for LTI systems in Section \ref{OpacityOutputCont}. 
To the best of our knowledge, there does not exist a notion of output controllability for DTNLSs of the form in Equation (\ref{NonLinSys}). 
It will be interesting to develop a framework that would relate $k$-ISO for nonlinear systems to (nonlinear) output controllability.  
\section{Conclusion and Future Directions} \label{Conclusion}

In this paper, we have proposed a framework for opacity for dynamical systems in which the state and input variables take continuous values. 
Prior work in this field studied opacity only for discrete event systems, in which states were discrete. 
We first presented a notion of opacity for single-adversary discrete-time, linear, time-invariant systems called $k$-initial state opacity. 
Necessary and sufficient conditions to establish $k$-ISO were presented in terms of sets of reachable states. 
Realizing that exact computation of these reachable sets is difficult, we presented a characterization of $k$-ISO using overapproximations and underapproximations of these sets. 
Further, we showed that opacity of a given DT-LTI system was equivalent to the output controllability of a system obeying the same dynamics, but with a different initial condition. 

We then proposed an extension to multiple-adversary systems, where we formulated more than one notion of decentralized opacity. 
These notions depended on whether there was a centralized coordinator or not, and the presence or absence of collusion among the adversaries. 
We established conditions for decentralized opacity in terms of reachable sets of states. 
In the case of colluding adversaries, we derived a condition for nonopacity in terms of the structure of the communication graph. 

Finally, we formulated notions of opacity for discrete-time nonlinear systems, and for the case when the assumptions on secret and nonsecret outputs being exactly equal was relaxed, called $\epsilon-k$-ISO. 
 
The results in this paper have been qualitative in nature. 
Future work will pursue means of quantifying opacity. 
A first step in this direction would be to investigate if there is a connection between the notion of $\epsilon-k$-ISO and differential privacy. 
We could leverage the results on $k$-ISO under unions and intersections in Section \ref{UandI} to quantify how opaque the secret is when one only has an arbitrary, finite number of overapproximations and underapproximations of the reachable sets of states. 
A third interesting scenario to consider is when the adversary might incur a cost in making an observation, and has to decide on opacity by incurring as low a cost as possible. 

The results in this paper, together with our work on opacity for switched linear systems in \cite{bhaskar2017switched}, provide a complete framework for opacity for continuous-state dynamical systems. 
Although our results have been for the discrete-time case, many of them will hold for continuous-time systems as well. 
This work complements the large body of literature on opacity for discrete event systems. 
It is our belief that the two frameworks can be used in conjunction to analyze opacity for more general CPSs. 
\bibliographystyle{ieeetr}
\bibliography{ThesisBib.bib}
\end{document}